\pdfoutput=1
\documentclass[conference]{IEEEtran}

\usepackage{times}
\usepackage{bbm}
\usepackage[cmex10]{amsmath}
\usepackage{graphicx}
\usepackage{amssymb}
\usepackage{amsfonts}
\usepackage{float}
\usepackage{stfloats}
\usepackage{cases}
\usepackage{algorithm}
\usepackage{algorithmic}
\usepackage{color}
\usepackage{array}
\usepackage{mdwmath}
\usepackage{mdwtab}
\usepackage{eqparbox}
\usepackage[tight,footnotesize]{subfigure}
\usepackage{multirow}
\usepackage{caption}
\usepackage{cite}
\usepackage{amsmath}
\usepackage{tabu}
\usepackage{tabularx, booktabs, array, rotating, multirow}
\usepackage{bm}
\usepackage{epstopdf}
\ifCLASSINFOpdf

\else

\fi

\begin{document}
\title{Opportunistic Scheduling for Network Coded Data in Wireless Multicast Networks}

\author{\IEEEauthorblockN{Nadieh Moghadam, Mohammad Mohebbi and Hongxiang Li}
\IEEEauthorblockA{Department of Electrical\\ and Computer Engineering\\
University of Louisville\\
Louisville, Kentucky 40292\\
Email: n0moha07@louisville.edu, m0mohe01@louisville.edu, h.li@louisville.edu}
}

\maketitle

\begin{abstract}
In this paper queue stability in a single-hop wireless multicast networks over erasure channels is analyzed. First, a queuing model consisting of several sub-queues is introduced. Under the queueing stability constraint, we adopt Lyapunov optimization model and define decision variables to derive a network coding based packet scheduling algorithm, which has significantly less complexity and shorter queue size compared with the existing solutions. Further, the proposed algorithm is modified to meet the requirements of time-critical data. Finally, the simulation results verify the effectiveness of our proposed algorithm.
\end{abstract}


\IEEEpeerreviewmaketitle

\section{Introduction}
\vspace{-1 mm}
Network Coding (NC) \cite{Ahlswede} is one of the most promising techniques to improve throughput in complex networks. In particular, it has been shown that NC can substantially improve network throughput and/or transmission delay over multicast networks. 
Most existing studies in multicast networks assume saturated transmitter queue that guarantees packet availability for transmission \cite{Medard2005,SVC,Qureshi,Amerimehr}, where the queue length is considered to be infinite. In this paper, realistic bounded queue is applied. For wireless networks, \cite{Hui} and \cite{Shrader} considered multi-hop and multi-source transmissions to find the maximum stable throughput. Furthermore, a stability policy is provided in \cite{XORBased} for a network coded unicast network. In delay sensitive applications, \cite{BufferAware} proposed a buffer-aware network coding method to reduce the delay caused by random packet arrivals. Additionally, stability properties are also evaluated in \cite{QueueStability} for broadcast/multicast erasure channels with NC, where the authors proposed a suboptimal virtual queue structure and provide detailed analysis for the case of two receivers. However, there is no closed form analysis for stability conditions and the proposed queueing model is quite complicated when the number of receivers is greater than two. In \cite{Sagduyu_2013} the authors analyzed the performance of a simple two-user broadcast channel in terms of stable throughput region.

Recently, the authors in \cite{Nadieh} proposed a wireless multicast virtual queue model for the special case of two receivers, where the transmitter has one sub-queue for each receiver. However, for wireless fading channels with a large number of receivers, it is likely that a packet fails at more than two receivers. Along the same line, an upper bound of the maximum input rate for a stable multicast system is derived in \cite{Nadieh2} for the case of three users. In \cite{Nadieh2} there is only one unique sub-queue, which stores those transmitted packets that are successfully received by at least one but not all receivers. Note that each packet in the sub-queue is associated with an \emph{index set} consisting of the receivers who have not received this packet; then the packets with disjoint index sets are combined and sent to all target receivers. Finally, in \cite{Nadieh3} linear programming with queueing stability constraints is used to obtain the optimal network coding scheduling policy. However, this solution is not scalable for a large number of users due to the prohibitive computational complexity.

In this paper, the queue structure is the same as \cite{Nadieh3}. However instead of selecting network coding scheduling schemes based on the probabilities resulted from the solution of a linear programming as it is presented in \cite{Nadieh3}, in this work with the aid of \emph{Lyapunov Optimization Model}, the scheduling schemes are chosen based on the minimization of a \emph{Decision Function}. Further, we expand our work by considering time-critical data where each packet expires after a predefined deadline and is then considered useless for any receiver and consequently dropped from the system. However, for packets with hard deadlines queue stability is trivial and irrelevant since expired packets are dropped in any case.
Specifically, our main contributions are summarized as follows:

\begin{itemize}
\item Based on network coding and Lyapunov Optimization Model, we propose new low-complexity multicast scheduling algorithm that not only meets the queue stability constraint but also minimizes the queue size.
\item To minimize the number of dropped packets resulted from hard deadline, we further derive a scheduling policy that simultaneously minimizes both dropped packets and queue size.
\end{itemize}

The rest of the paper is organized as follows. The system model is presented in Section II. A Lyapunov based scheduling along with stability analysis are provided in section III followed by the modified analysis for time-critical data in section IV. In Section V, the performance of the proposed scheduling is validated through simulation. Finally, a conclusion is drawn in Section VI.
\vspace{-3 mm}
\section{System Model}
We consider a one-hop wireless multicast system in which one transmitter multicasts data packets to $N$ users over erasure channels. The packets are delivered to the source for transmission over erasure channels according to a stationary process with arrival rate $\lambda$. We consider random and unreliable multicast channels where each packet transmission fails at receivers independently with packet error rate $\epsilon_{i}$ ($i = 1, 2, ..., N$), i.e., the packet is received successfully by receiver $i$ with probability $\gamma_i=1-\epsilon_i$. We assume the system is time slotted. At the beginning of every time slot, each receiver sends a one bit feedback to the transmitter indicating whether the previously sent packet has been successfully received or not.

A queue is considered stable if the arrival rate is less than the service rate. For queue stability analysis, we consider stationary operation when the queue distribution reaches a steady state. Let $\mu$ be the service rate of the source queue, the stability condition is given by $\lambda/\mu<1$.
Further, for a time-critical system it is assumed that each packet is associated with a packet delivery deadline after which the packet is considered useless.

In this paper we explore network coding based multicast scheduling schemes. We adopt the same queuing model as in \cite{Nadieh3}, but the scheduling policy is totally different from that of \cite{Nadieh3}.
Specifically, the queue system at the transmitter consists of $M$ sub-queues with $M=2^N-1$. Each sub-queue $q_{i}$ ($i=0, 1,...,M-1$) is associated with a unique user index set $I_i$, indicating its intended users (i.e., all users that are still expecting packets from $q_{i}$). Note that all the newly arrived packets are first stored in $q_{0}$ whose user index set is $I_0=\{1,...,N\}$. 
In each time slot, according to certain network coding scheme, packets from some different sub-queues (i.e., NC participating packets) are combined into a network coded packet and transmitted. It is worth noting that, to simultaneously minimize the queue size and benefit the maximum number of users, the index sets of the selected sub-queues are disjoint and their union forms a complete user set. After each transmission, based on the receiver feedback, a participating packet may: (1) stay in the same sub-queue if none of its intended users received the network coded packet; or (2) leaves the queue system if all of its intended users received the network coded packet; or (3) moves to another sub-queue whose user index set matches the participating packet's new intended users.

The queue structure for the case of three receivers is provided in Table 1 with five network coding scheduling schemes. For example, $S_0$ means to transmit the head-of-line packet from $q_{0}$; $S_1$ means to combines the head-of-line packets from $q_{4}$, $q_{5}$ and $q_{6}$ for transmission. Accordingly, Fig. 1 illustrates the \emph{Data-Flow} model \cite{Nadieh3}, where packets enter $q_0$ by rate $\lambda$ and eventually are received by all users and leave the queueing system.

\setlength{\belowcaptionskip}{-15pt}

\begin{table}[]
\centering
\caption{Network coding scheduling policies for a system with 3 receivers.}
\label{my-label}
\def\arraystretch{1.7}
\setlength{\tabcolsep}{0.4em}
\begin{tabular}{c|c|c|c|c|c|c|c|c|}
\cline{2-9}
 & \textbf{Sub-queue} & $q_0$ & $q_1$ & $q_2$ & $q_3$ & $q_4$ & $q_5$ & $q_6$ \\ \cline{2-9}
\textbf{} & \textbf{Index Set} & \{1,2,3\} & \{1,2\} & \{2,3\} & \{1,3\} & \{1\} & \{2\} & \{3\} \\ \hline
\multicolumn{1}{|c|}{\multirow{5}{*}{\begin{turn}{-270}\textbf{NC Scheduling}\end{turn}}} & $S_0$ & 1 & 0 & 0 & 0 & 0 & 0 & 0 \\ \cline{2-9}
\multicolumn{1}{|c|}{} & $S_1$ & 0 & 0 & 0 & 0 & 1 & 1 & 1 \\ \cline{2-9}
\multicolumn{1}{|c|}{} & $S_2$ & 0 & 1 & 0 & 0 & 0 & 0 & 1 \\ \cline{2-9}
\multicolumn{1}{|c|}{} & $S_3$ & 0 & 0 & 1 & 0 & 1 & 0 & 0 \\ \cline{2-9}
\multicolumn{1}{|c|}{} & $S_4$ & 0 & 0 & 0 & 1 & 0 & 1 & 0 \\ \hline
\end{tabular}
\end{table}
\begin{figure}[]
\centering
\includegraphics[width=3.8in]{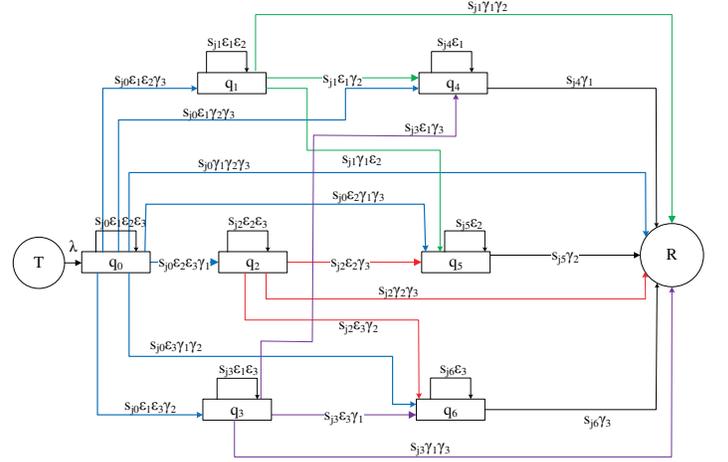}
\caption{Architecture of the Data-Flow model.}
\end{figure}

Our first objective is to find the optimal network coding scheduling scheme to keep the queue system stable.
To solve this problem, \cite{Nadieh3} proposed Linear Programming based Scheduling method (LPS) and derived the following capacity region
\vspace{-3 mm}
\begin{equation}
\Lambda=\{\lambda:\lambda<1-max(\epsilon_1,\epsilon_2,...,\epsilon_N)\},
\end{equation}

However, the LPS approach is not suitable for dynamic fading channels, where the packet error rates vary from time slot to time slot so that the linear programming problem must be solved at the beginning of every time slot. To reduce the computational complexity, in this paper we apply \emph{Lyapunov Optimization Model} by defining Decision Variables (DV) and derive new scheduling scheme that keeps the system \emph{strongly stable}. Note that a \emph{strongly stable} queueing system  is defined in \cite{Neely} as follows: 

\emph{Definition 1:} A discrete time process $q(t)$ is \emph{strongly stable} if
\begin{equation}\label{eq:Def1}
  \limsup_{t\to\infty} \frac{1}{t}\sum_{\tau=0}^{t-1}\mathbb{E}\{|q(\tau)|\}< \infty
\end{equation}
where $|q(\tau)|$ is the cardinality of $q(\tau)$.

An extension of this definition for a system with $M$ sub-queues can be presented as
\begin{equation}\label{eq:Def2}
  \limsup_{t\to\infty} \frac{1}{t}\sum_{\tau=0}^{t-1}\sum_{i=0}^{M-1}\mathbb{E}\{|q_i(\tau)|\}\leq \infty
\end{equation}


\section{Lyapunov Based Network Coding Scheme}
Using Lyapunov optimization model, for any input rate in the capacity region [see Eq. (1)], we aim to obtain a scheduling scheme that not only guarantees queueing stability but also minimize the sub-queue lengths. In the rest of this paper, this new approach is called \emph{Lyapunov based Scheduling (LyS)}. 

Defining sub-queue length vector $\bm{Q}=[Q_0, Q_1, ..., Q_{M-1}]$ with $Q_i$ being the queue length (backlog) for sub-queue, $q_i$, the queueing dynamic can be written as
\begin{equation}\label{eq:q_dynamic}
Q_i(t+1) =max[Q_i(t)-D_i(t),0]+A_i(t) \hspace{5mm} \forall i= 0,1,...,M-1
\end{equation}
where $A_i$ and $D_i$ are respectively the total arrival rate and departure rate for each sub-queue.

The \emph{Lyapunov function} \cite{Neely} for a queueing system with $M$ sub-queues is defined as
\vspace{-3 mm}
\begin{equation}\label{eq:Ly_Func}
L(t) = \frac{1}{2}\sum_{i=0}^{M-1}Q_i^2(t)
\end{equation}

Accordingly, \emph{Lyapunov Drift} is defined in (\ref{eq:Ly_Drift}) as the change in Lyapunov function from one time slot to the next,
\begin{equation}\label{eq:Ly_Drift}
  \Delta L(t)=L(t+1)-L(t)=\frac{1}{2}\sum_{i=0}^{M-1}[Q_i^2(t+1)-Q_i^2(t)]
\end{equation}

Plugging (\ref{eq:q_dynamic}) into (\ref{eq:Ly_Drift}), we have
\begin{equation}\label{eq:Ly_uneq}
   \Delta L(t)\leq \sum_{i=0}^{M-1}\frac{A_i^2(t)+D_i^2(t)}{2}+\sum_{i=0}^{M-1}Q_i(t)[A_i(t)-D_i(t)]
\end{equation}

Then, we define \emph{conditional Lyapunov drift} as
\begin{equation}\label{eq:C_Ly_Drift}
  \Delta L_c(t)=\mathbb{E}\{(L(t+1)-L(t))|Q(t)\}
\end{equation}
where the expectation depends on the randomness of the channel.
Similarly, using (\ref{eq:q_dynamic}) we have
\begin{multline}\label{eq:C_Ly_uneq}
   \Delta L_c(t)\leq \mathbb{E}\{\sum_{i=0}^{M-1}\frac{A_i^2(t)+D_i^2(t)}{2}|Q(t)+\\
   \sum_{i=0}^{M-1}Q_i(t)[A_i(t)-D_i(t)]|Q(t)\}
\end{multline}

Since $A_i(t)$ and $D_i(t)$ are bounded, there is a finite constant, $B$, such that
\begin{equation}
\mathbb{E}\{\sum_{i=0}^{M-1}\frac{A_i^2(t)+D_i^2(t)}{2}|Q(t)\}\leq B
\end{equation}
\vspace{-2 mm}
Thus, we get
\vspace{-3 mm}
\begin{multline}\label{eq:C2_Ly_uneq}
   \Delta L_c(t)\leq B + \mathbb{E}\{
   \sum_{i=0}^{M-1}Q_i(t)[A_i(t)-D_i(t)]|Q(t)\}
\end{multline}
We further define decision function as
\begin{equation}\label{eq:min}
  F(t,S_j)\triangleq \mathbb{E}\{\sum_{i=0}^{M-1}Q_i(t)[A_i(t)-D_i(t)]|Q(t)\},
\end{equation}

It is obvious that this expectation depends on the selected schedule in each time slot; therefore, the decision function, $F(t,S_j)$, is denoted as a function of $S_j$.

In order to prove that the system is strongly stable, according to (\ref{eq:Def2}) we need to show
\vspace{-3 mm}
\begin{equation}\label{}
  \limsup_{t\to\infty} \frac{1}{t}\sum_{\tau=0}^{t-1}\sum_{i=0}^{M-1}\mathbb{E}\{Q_i(\tau)\}\leq \infty
\end{equation}
\vspace{-3 mm}

In the proceeding of the analysis, we show that a scheduling scheme which minimize $F(t,S_j)$, results in a strongly stable system.

\emph{Theorem 1:} Assuming the input rate is in the capacity region, with the NC scheduling scheme, $S^*_j$, defined as
\vspace{-1 mm}
\begin{equation}\label{}
  S^*_j \triangleq \underset{S_j}{\operatorname{argmin}} (F(t,S_j))
\end{equation}
\vspace{-3 mm}

the queueing system is strongly stable.

\emph{Proof:}

It is obvious that if $A_i \leq D_i$ for sub-queue, $q_i$, then that sub-queue is stable. For the stability for a system with $M$ sub-queue it is necessary that all sub-queues be stable.
In this case we define the difference, $\delta \geq 0$, such that for each sub-queue we have:
$A_i+\delta \leq D_i, \forall i$.
In order to proceed the stability proof, we need to define the following LP which is independent from the sub-queue lengths. This LP provides a scheduling scheme that keeps the system in the stability region while the difference between arrival rate and the departure rate is maximized.
\vspace{-3 mm}
\begin{flalign}\label{LP2_1}
\begin{split}
&\max~~  \delta  \\
s.t.  \\
&\mathbb{E}\{A_i\}+ \delta \leq \mathbb{E}\{D_i\} \\
&\forall i= 0,1,...,M-1
\end{split}
\end{flalign}
\vspace{-3 mm}

For a three user scenario, using the Data-Flow model in Fig. 1, the LP optimization problem in (\ref{LP2_1}) becomes
\begin{flalign}\label{eq:LP2}
\begin{split}
&\max~~  \delta  \\
&s.t.  \\
c_0:&\lambda + \delta \leq p_0(\epsilon_1\epsilon_2\gamma_3+\epsilon_1\gamma_2\gamma_3+\gamma_1\gamma_2\gamma_3+\epsilon_2\gamma_1\gamma_3+\\
&\epsilon_2\epsilon_3\gamma_1+\gamma_1\gamma_2\epsilon_3+\epsilon_1\epsilon_3\gamma_2);\\
c_1:&p_0\epsilon_1\epsilon_2\gamma_3 + \delta \leq p_2(\gamma_1\gamma_2+\epsilon_1\gamma_2+\gamma_1\epsilon_2);\\
c_2:&p_0\epsilon_2\epsilon_3\gamma_1 + \delta \leq p_3(\epsilon_2\gamma_3+\gamma_2\gamma_3+\epsilon_3\gamma_2);\\
c_3:&p_0\epsilon_1\epsilon_3\gamma_2 + \delta \leq p_4(\gamma_1\gamma_3+\epsilon_1\gamma_3+\gamma_1\epsilon_3);\\
c_4:&p_2\epsilon_1\gamma_2+p_0\epsilon_1\gamma_2\gamma_3+p_4\epsilon_1\gamma_3 + \delta \leq (p_1+p_3)\gamma_1;\\
c_5:&p_2\gamma_1\epsilon_2+p_0\epsilon_2\gamma_1\gamma_3+p_3\epsilon_2\gamma_3 + \delta \leq (p_1+p_4)\gamma_2;\\
c_6:&p_3\epsilon_3\gamma_2+p_0\epsilon_3\gamma_1\gamma_2+p_4\epsilon_3\gamma_1 + \delta \leq (p_1+p_2)\gamma_3;\\
&\sum_{i=0}^{4} p_j  =  1; \\
& p_j \geq  0~~ \forall j
\end{split}
\end{flalign}
where $\{p_0,..., p_4\}$ are the optimization variables representing optimal probabilities based on which a scheduling scheme is chosen; constraint $c_i$ guarantees the stability of sub-queue $i$.

Denote $S^{\dag}_j$ as the optimal scheduling solution to (\ref{LP2_1}), we have
\vspace{-3 mm}
\begin{equation}\label{}
 \delta_{max} \leq \mathbb{E}\{D_i^{\dag}\}-\mathbb{E}\{A_i^{\dag}\},\quad \forall i= 0,2,...,M-1
\end{equation}
Since $S^*_j$ minimizes $F(t,S^*_j)$, we have $F(t,S^*_j)\leq
F(t,S^{\dag}_j)$. Therefore,
\vspace{-4 mm}
\begin{multline}\label{18}
  \Delta L^*_c(t)\leq B +\mathbb{E}\{\sum_{i=0}^{M-1}Q_i(t)[A^*_i(t)-D^*_i(t)]|Q(t)\} \leq \\
  B +\mathbb{E}\{\sum_{i=0}^{M-1}Q_i(t)[A^{\dag}_i(t)-D^{\dag}_i(t)]|Q(t)\}
\end{multline}

Because the scheduling scheme in (\ref{LP2_1}) is independent of the queue lengths, we have
\vspace{-3 mm}
\begin{multline}\label{19}
\mathbb{E}\{\sum_{i=0}^{M-1}Q_i(t)[A^{\dag}_i(t)-D^{\dag}_i(t)]|Q(t)\} = \\
\mathbb{E}\{\sum_{i=0}^{M-1}Q_i(t)[A^{\dag}_i(t)-D^{\dag}_i(t)]\} \leq -\delta_{max}\sum_{i=0}^{M-1}\mathbb{E}\{Q_i(t)\}
\end{multline}
\vspace{-4 mm}
From (\ref{18}) and (\ref{19}), we have
\begin{equation}\label{20}
  \Delta L^*_c(t)\leq B -\delta_{max}\sum_{i=0}^{M-1}\mathbb{E}\{Q_i(t)\}
\end{equation}
Taking expectation on both sides of (\ref{20}) yields
\begin{multline}\label{21}
  \mathbb{E}\{\Delta L^*_c(t)\}= \mathbb{E}\{ \mathbb{E}\{L^*_c(t+1)\}-\mathbb{E}\{L^*_c(t)\}\}\\
  \leq B -\delta_{max}\sum_{i=1}^{M-1}\mathbb{E}\{Q_i(t)\}
\end{multline}
Summing over $t\in\{0,1,...,T-1\}$, we have
\begin{equation}\label{}
  \mathbb{E}\{L^*_c(T)\}-\mathbb{E}\{L^*_c(0)\}\leq BT-\delta_{max}\sum_{t=0}^{T-1}\sum_{i=0}^{M-1}\mathbb{E}\{Q_i(t)\}
\end{equation}
Considering the fact that $\mathbb{E}\{L^*_c(T)\}\geq0$, we have
\begin{equation}\label{23}
  \frac{1}{T}\sum_{t=0}^{T-1}\sum_{i=0}^{M-1}\mathbb{E}\{Q_i(t)\}\leq\frac{B}{\delta_{max}}+\frac{\mathbb{E}\{L_c(0)\}}{T\delta_{max}}
\end{equation}
where the negative term is eliminated from the right hand side of (\ref{23}).
Taking the lim sup of both sides of (\ref{23}) gives

\begin{equation}\label{eq:Ly_last}
  \limsup_{T\to\infty} \frac{1}{T}\sum_{t=0}^{T-1}\sum_{i=0}^{M-1}\mathbb{E}\{Q_i(t)\}\leq\frac{B}{\delta_{max}}
\end{equation}
Eq. (\ref{eq:Ly_last}) shows that the system is strongly stable 
and the total average queue length is less than or equal to $B/\delta_{max}$. $\blacksquare$


\section{Lyapunov Model for Time-Critical Data}
So far, the main goal was queue size minimization which leads to queue stability. Packet deadline was not considered in the scheduling schemes. Minimizing Lyapunov drift as it is done in the previous section may lead to existence of extremely aged packets in some sub-queues which are consequently dropped. Nevertheless, in order to decrease the number of dropped packets, we need to take into account the age of the packets in each sub-queue, i.e. oldest packets (which are closer to their deadline) have higher priority for transmission. In this regard we define \emph{drift-plus-penalty} \cite{Neely} as $\Delta L(t) + \beta a_o(t)$; where $\beta \geq0$ is a parameter called \emph{importance weight} that balances queue size minimization and age minimization and $a_o(t)$ is the age of the oldest packets in the involved sub-queues of the scheduling scheme in time slot $t$.

We have already calculated a bound for $\Delta L_c(t)$ in (\ref{eq:C2_Ly_uneq}). Adding $\beta a_o(t)$ to both sides of (\ref{eq:C2_Ly_uneq}) yields a bound on the drift-plus-penalty as
\begin{multline}\label{eq:drift}
   \Delta L_c(t) + \beta a_o(t) \leq B + \beta a_o(t) + \\
   \mathbb{E}\{
   \sum_{i=0}^{M-1}Q_i(t)[A_i(t)-D_i(t)]|Q(t)\}
\end{multline}

So the new decision function, $F'(t,S_j)$, is defined as
\begin{equation}\label{Eq:Fp}
  F'(t,S_j) = \beta a_o(t) + \mathbb{E}\{
   \sum_{i=0}^{M-1}Q_i(t)[A_i(t)-D_i(t)]|Q(t)\}
\end{equation}
which is a function of NC scheduling scheme $S_j$.

A scheduling scheme which minimize $F'(t,S_j)$, makes a tradeoff between queue backlog and packet dropping.
This optimal NC scheduling scheme, $S'^*_j$, is defined as

\begin{equation}\label{}
  S'^*_j \triangleq \underset{S_j}{\operatorname{argmin}} (F'(t,S_j))
\end{equation}


In the next section we show that using this scheduling scheme the number of dropped packets is decreased.

\section{Simulation Results}
In this section, we use simulation to validate the performance of LyS model. For simplicity we consider a three user scenario in our simulations. Opportunistically minimization of expectation of (\ref{eq:min}) is through finding the minimum of the decision variables defined as
\vspace{-1 mm}
\begin{equation}\label{sim1}
  DV_j = \sum_{i\in V_j} Q_i(t)(A_i(t)-D_i(t))
\end{equation}
where $V_j=\{$sub-queues involved in schedule $j\}$.

In every time slot the minimum $DV_j$ is selected as the scheduling scheme.
For a time-critical system with packet deadline (\ref{sim1}) becomes
\begin{equation}\label{sim2}
  DV'_j = \beta a_o(t) + \sum_{i\in V_j} Q_i(t)(A_i(t)-D_i(t))
\end{equation}
\begin{figure}[t]
\centering
\includegraphics[width=3in]{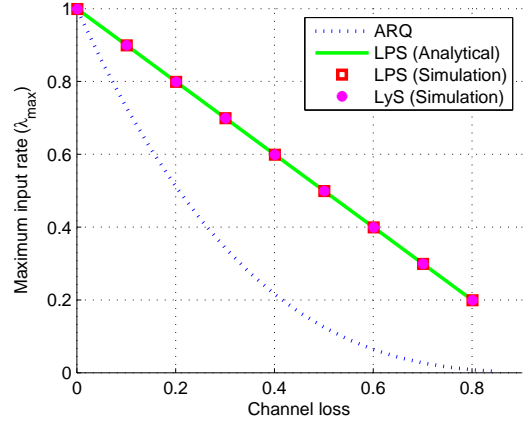}
\caption{Maximum input rate comparison with 3 receivers.}\label{Comparison}
\end{figure}

Fig. \ref{Comparison} illustrates the maximum input rate, $\lambda_{max}$, versus $\epsilon$ when the queuing system is stable for different methods including ARQ, LPS and LyS. For all methods, as expected, the maximum allowable input rate decreases with $\epsilon$. As it can be seen the performance of LyS matches that of LPS which is proven to be optimal scheduling scheme in \cite{Nadieh3}.
More importantly, the LPS and LyS models outperforms the traditional ARQ.
In addition to queueing stability, we study the backlog of sub-queues. As it can be inferred from (\ref{sim1}), the decision variables in LyS are based on queue length; this fact results in a shorter backlog in LyS compared to LPS. For instance with the input rate equals to \%70 and variable channel loss of $\mathbb{E}\{\epsilon_i\} = \%20$ for $i= 1,2,3$, the average $Q_T$ for LPS and LyS methods are 44.8 and 3.9 respectively.

\begin{figure}[]
\centering
\includegraphics[width=2.7in]{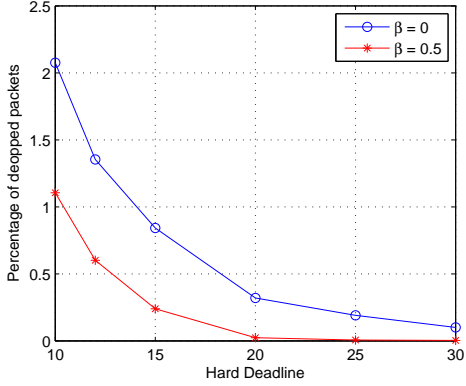}
\caption{Percentage of dropped packets versus hard deadline.}\label{4}
\end{figure}

\begin{figure}[]
\centering
\includegraphics[width=2.7in]{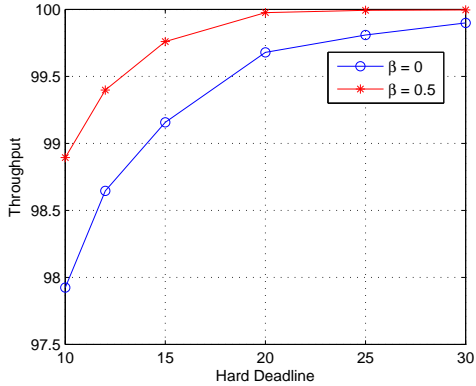}
\caption{Percentage of throughput versus hard deadline.}\label{5}
\end{figure}

Fig. \ref{4} and Fig. \ref{5} depict the percentage of dropped packets and throughput versus different values of hard deadline for $\beta=0$ and $\beta=0.5$. It is clear that as the hard deadline increases, we have less dropped packets and higher throughput.
Moreover, with $\beta=0.5$ where we give transmission priority to the sub-queues with aged packets, we have less percentage of dropped packets and consequently higher throughput.
Be noted that when $\beta = 0$, (\ref{sim2}) is reduced to (\ref{sim1}) in which only the queue size is minimized.

\begin{figure}[]
\centering
\includegraphics[width=2.7in]{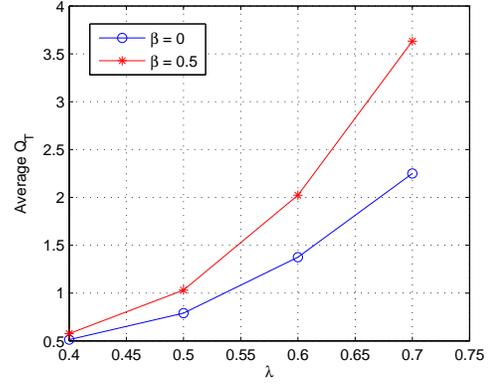}
\caption{Average $Q_T$ versus $\lambda$.}\label{meanQT}
\end{figure}

Fig. (\ref{meanQT}) illustrates $Q_T$ in LyS model versus $\lambda$ for different values of $\beta$ when hard deadline is equal to 10. As it is expected $Q_T$ is slightly higher when we have nonzero $\beta$. As previously mentioned the proposed scheduling provides a tradeoff between queue length and dropped packets.

\section{Conclusion}
\vspace{-1 mm}
In this paper by applying Lyapunov optimization model we develop the optimal scheduling scheme such that for a given input rate in the capacity region, this scheduling always keeps the system stable. This method has better performance over LP based scheduling in terms of complexity and queue length. Furthermore, using drift-plus-penalty we introduce a tradeoff between queue length and dropped packets minimization to develop a scheduling scheme for time-critical data.

\end{document}